\documentclass[prd,aps,a4paper,superscriptaddress,twocolumn,nofootinbib]{revtex4}
\usepackage{graphicx}
\usepackage{color}
\usepackage{xcolor}
\usepackage{dcolumn}
\usepackage{bm}
\usepackage{slashed}
\usepackage{amsmath}
\usepackage{latexsym}
\usepackage{amssymb}
\usepackage{mathrsfs}
\usepackage{amsfonts}
\usepackage{xspace}
\usepackage{ulem}

\allowdisplaybreaks

\begin{document}

\title{Solving Hamiltonian Constraint Equation with Physics-Informed Neural Networks}
\author{Yu-Chen Zhou} %\email{zhouyuchen21@mails.ucas.ac.cn}
\affiliation{School of Fundamental Physics and Mathematical Sciences, Hangzhou Institute for Advanced Study, UCAS, Hangzhou 310024, China}
\affiliation{Department of Artificial Intelligence, School of Engineering, Westlake University, Hangzhou 310030, Zhejiang Province, China}
\affiliation{School of Physical Sciences, University of Chinese Academy of Sciences, Beijing 100049, China}
\affiliation{International Centre for Theoretical Physics Asia-Pacific, University of Chinese Academy of Sciences, 100190 Beijing, China}
\author{Hao Ma} %\email{mahao@westlake.edu.cn}
\affiliation{Department of Astronomy, Westlake University, Hangzhou 310030, Zhejiang Province, China}
\author{Zhoujian Cao} \email[Zhoujian Cao: ]{zjcao@amt.ac.cn}
\affiliation{Institute for Frontiers in Astronomy and Astrophysics, Beijing Normal University, Beijing 102206, China}
\affiliation{School of Physics and Astronomy, Beijing Normal University, Beijing 100875, China}
\affiliation{School of Fundamental Physics and Mathematical Sciences, Hangzhou Institute for Advanced Study, UCAS, Hangzhou 310024, China}
\author{Tailin Wu} \email[Tailin Wu: ]{wutailin@westlake.edu.cn}
\affiliation{Department of Artificial Intelligence, School of Engineering, Westlake University, Hangzhou 310030, Zhejiang Province, China}
\author{Hong-Bo Jin} \email[Hong-Bo Jin: ]{hbjin@bao.ac.cn}
\affiliation{National Astronomical Observatories, Chinese Academy of Sciences, Beijing 100101, China}
\affiliation{School of Fundamental Physics and Mathematical Sciences, Hangzhou Institute for Advanced Study, UCAS, Hangzhou 310024, China}
\author{Xuefeng Feng} %\email{fengxuefeng@simis.cn}
\affiliation{Fudan Center for Mathematics and Interdisciplinary Study, Fudan University, Shanghai, 200433, China}
\affiliation{Shanghai Institute for Mathematics and Interdisciplinary Sciences (SIMIS), Shanghai, 200433, China}
\author{Shuanglin Huang} %\email{hslynn@amss.ac.cn}
\affiliation{College of Mathematics and Physics, China Three Gorges University, Yichang, 443002, China}
\author{Zhi-Chao Zhao} %\email{zhaozc@cau.edu.cn}
\affiliation{Department of Applied Physics, College of Science, China Agricultural University, Qinghua East Road, Beijing 100083, the People's Republic of China}
\author{Yue-Liang Wu} %\email{ylwu@itp.ac.cn}
\affiliation{School of Physical Sciences, University of Chinese Academy of Sciences, Beijing 100049, China}
\affiliation{International Centre for Theoretical Physics Asia-Pacific, University of Chinese Academy of Sciences, 100190 Beijing, China}
\affiliation{School of Fundamental Physics and Mathematical Sciences, Hangzhou Institute for Advanced Study, UCAS, Hangzhou 310024, China}
\affiliation{Institute of Theoretical Physics, Chinese Academy of Sciences, Beijing 100190, China}

\begin{abstract}
Numerical relativity (NR), solving Einstein equation numerically, plays an important role in source modelling for gravitational wave astronomy. Traditional methods for NR including finite difference method, spectral method and finite element method have been well developed. But newly developed neural network methods for partial differential equations (PDE) have not been well studied yet for NR. We present a Physics-Informed Neural Network (PINN) method to solve the Hamiltonian constraint equation for binary black hole (BBH) initial data in NR. This equation is a highly non-linear elliptic PDE, posing significant challenges for conventional PINN approaches. To overcome these difficulties, we introduce a set of new techniques. We show that our PINN together with these techniques can successfully solve the Hamiltonian constraint equation for generic BBH systems. Validation against the traditional results demonstrates the high accuracy and robustness of our method, revealing the immense potential of constructing a PINN-based initial data solution to all BBH systems for NR.
\end{abstract}

\maketitle

\section{Introduction}\label{s1}
Since the breakthrough of GW150914 in 2015, the gravitational wave astronomy has developed very rapidly. Source modelling played an important role in gravitational wave data processing. In order to model gravitational wave sources, numerical relativity (NR) is an indispensable tool.

The compact binary object coalescence (CBC) is among the most important gravitational wave sources. For CBCs, the existing NR techniques solve the corresponding Einstein equation one CBC by one CBC. But physically we can believe all CBC gravitational waveforms must admit some common property. Like post-Newtonian approximation can present us a waveform formula which depends on the CBC parameters, it is useful to construct a numerical solution depending on CBC parameters for all CBCs. This task seems impossible for traditional NR techniques. In contrast, neural network method is promising for this task.

In recent years, solving differential equations via machine learning methods has become a popular direction. A particularly powerful architecture in this domain is the Physics-Informed Neural Network (PINN) method \cite{RAISSI2019686}, which incorporates the physical laws as constraints by inserting differential equations into the loss function. The solution of the equations can then be obtained by minimizing the loss function. One of the distinct advantages of PINN method is its mesh-free nature, which offers more flexibility compared to many classical numerical methods.

PINNs have been successfully applied to solve partial differential equations (PDE) in many aspects, including simulations of fluid dynamics \cite{cai2021physicsinformedneuralnetworkspinns,RAO2020207,Farkane2023EPINNNSEEP,donnelly2024physics,HU2024107453,zhou2025hybrid,roy2024finite,jangir2026parameterized,pal2025solving,eivazi2022physics,wang2025discovery,coutinho2023physics}, solving Schr\"odinger equations \cite{shah2022physics,jin2022physics,sarkar2025physics,harcombe2023physics,mattheakis2022first}, black hole quasinormal modes (QNMs) \cite{Luna:2022rql, Luna:2024spo} and pulsar equations \cite{Stefanou:2023jxk}. Regarding Einstein equations, the application of neural network method is much fewer. To our knowledge, such applications only include the simulation of gravitational collapse \cite{Ferrer-Sanchez:2025cns}, and the solution of the Hamiltonian constraint equation in a conformally flat, time-symmetric vacuum system \cite{10.5120/ijca2026926359}.

NR reformulates the Einstein field equations into a 3+1 Cauchy initial value problem. With this formulation, we can solve the evolution processes of many astrophysical systems, such as binary black holes (BBHs) mergers \cite{Pretorius:2005gq, Baker:2005vv, Campanelli:2005dd}, coalescence of binary neutron stars (BNSs) \cite{Shibata:1999hn, Shibata:1999wm} and Black hole-neutron star (BHNS) binaries \cite{Shibata:2006bs, Etienne:2008re}. Numerical simulations of these processes are of great significance to the gravitational wave astronomy, because high precision gravitational wave templates can be extracted from the simulations, which are necessary for the gravitational wave detections and parameter estimations \cite{LIGOScientific:2016vbw}.

In order to simulate such systems, one must first construct suitable initial data. For BBH systems, we need to solve the Hamiltonian constraint equations \cite{Cook:2000vr, Pfeiffer:2004nc}. The Hamiltonian constraint equation is a highly nonlinear elliptic equation \cite{Dain:2001ry} (Lichnerowicz equation) and can only be solved accurately through numerical methods. Commonly used methods are finite difference methods \cite{Yo:2004ng} and Spectral methods \cite{Ansorg:2004ds}.

In this work, we solve the Hamiltonian constraint equation in BBH systems by PINN methods for the first time. This highly nonlinear equation is typically difficult to solve with PINN. To address this, we incorporate an approximate analytical solution as a guiding component of the network. We also propose a two-stage training method to decrease the difficulties for the network. With these approaches, we manage to solve this equation successfully. In addition, to better resolve the local features near the punctures, we formulate a composite optimization framework that combines the $L_2$ loss with a $\text{soft-}L_\infty$ loss, together with a loss balancing strategy to keep different loss terms at comparable numerical scales during training. Furthermore, we perform several comparisons between the PINN predicted solutions and the reference solutions solved by \texttt{TwoPunctures} \cite{Ansorg:2004ds}. We also compare the results under different training strategies, demonstrating that our method can achieve better solutions.

The structure of this paper is as follows: In Section \ref{s2} we introduce the methodology, including the Hamiltonian constraint equation, the PINN ansatz, the loss-balancing strategy, and the training procedure. In Section \ref{s3} we present the ablation studies. In Section \ref{s4} we report the results for different binary black hole configurations and compare them with the \texttt{TwoPunctures} solutions. In Section \ref{s5} we summarize the main conclusions and outline future work.

\section{Methodology}\label{s2}

\subsection{Hamiltonian constraint equation}\label{s2a}

Based on the assumption that the three metric is conformally flat $\gamma_{ij}=\psi^{4}f_{ij}$ and the spacial slice is maximal $K=0$ \cite{Cook:2000vr, Pfeiffer:2002iy}, the Hamiltonian constraint equation can be written as
\begin{equation}
    \label{eq:psiequation}
    \Delta\psi + \frac{1}{8}\psi^{-7}\bar{K}_{ij}\bar{K}^{ij}=0,
\end{equation}
where $\bar{K}_{ij}=\psi^2 K_{ij}$ can be given by the Bowen-York solution \cite{PhysRevD.21.2047} to the momentum constraints, which can be written in the form

\begin{align}
\label{eq:Kij}
\bar{K}_{ij} = &\sum_n \Bigg[
 \frac{3}{2r_n^3} \left[ x_n^i P_n^j + x_n^j P_n^i - \left( \delta^{ij} - \frac{x_n^i x_n^j}{r_n^2} \right) P_k^n x_n^k \right] \nonumber\\
& + \frac{3}{r_n^5} \left( \epsilon^{ik}{}_{l} S_k^n x_n^l x_n^j + \epsilon^{jk}{}_{l} S_k^n x_n^l x_n^i \right) \Bigg],
\end{align}
where $r_n = \|\vec{x} - \vec{x}_n\|$, the index $n=1,2$ labels the two punctures, $P_i$ and $S_i$ represent the linear momentum and spin of the black holes, respectively. We use $\vec{x}$ and $\vec{x}_n$ to denote the field point and the position of the two punctures.

The conformal factor $\psi$ can be decomposed into the sum of a singular part and a finite correction term $u$ \cite{Brandt:1997tf}:
\begin{equation}
\label{eq:psi}
\psi = 1+\sum_n\frac{m_n}{2r_n}+u.
\end{equation}
The correction term $u$ satisfies the condition $u \rightarrow 0$ as $r \rightarrow \infty$. Substituting Eq. (\ref{eq:psi}) into the Hamiltonian constraint, we obtain the equation for the correction term $u$
\begin{equation}
\Delta u + \frac{1}{8}\psi^{-7}\bar{K}_{ij}\bar{K}^{ij}=0.\label{eq:uequation}
\end{equation}
In the current paper we will design a PINN method to solve the above equation.

\subsection{Architecture of our PINN method for Hamiltonian constraint equations}
\label{s2b}
\subsubsection{Guided hard-enforcement ansatz}
Since Eq.~(\ref{eq:uequation}) is a highly non-linear elliptic partial differential equation, it is very difficult to learn the solution $u$ directly by PINN. To reduce the difficulty of network training, we introduce the analytical formula for $u$ from \cite{Lousto:2007rj} as a guidance and assume that the solution $u_\theta$ learned by the neural network takes the following ansatz:
\begin{equation}
\label{eq:ansatz}
u_\theta(x) = \kappa\,u_{\text{g}}(x)\bigl[1 + cW(x)\tanh(h_\theta(x))\bigr],
\end{equation}
where $c$ is a tunable hyperparameter that controls the overall magnitude of the correction term, it adjusts how strongly the network-modulated correction contributes relative to the guiding component. $u_{\text{g}}$ is the approximate analytical solution to Eq.~(\ref{eq:uequation}), given by
\begin{equation}
\label{eq:guidance}
u_{\text{g}}=u_P+u_J+u_c,
\end{equation}
the definitions of $u_P$, $u_J$, and $u_c$ are given in \cite{Lousto:2007rj}. In the ansatz~(\ref{eq:ansatz}), we use the parameter $\kappa$ to adjust the global scale and the parameter $c$ to control the adjustment scale. Furthermore, we design a window function based on the magnitude of the guiding solution
\begin{equation}
\label{eq:wguidance}
W(x) = \frac{u_{\text{g}}(x)-u_{\min}}{u_{\max}-u_{\min}},
\end{equation}
where $u_\text{min}$ and $u_\text{max}$ denote the minimum and maximum values of the guiding function $u_{\text{g}}(x)$ respectively. Based on the ansatz~(\ref{eq:ansatz}), we use PINN to solve $h_\theta(x)$.

The original Hamiltonian constraint equation (\ref{eq:uequation}) is defined on whole $R^3$. Numerically we consider a spherical domain $\Omega$. On the boundary of this domain $\Omega$ (at $r=R_{\max}$) we set a Robin boundary condition \cite{PhysRevD.91.044033}
\begin{eqnarray}
\vec{n}\cdot\nabla u+\frac{1}{r}\frac{\partial r}{\partial n} u = 0\text{ at } \partial\Omega,\label{robin}
\end{eqnarray}
with $r=\sqrt{x^2+y^2+z^2}$ and $\vec{n}$ is the out normal vector of $\partial\Omega$. In the current work we set $R_{\max}=30$.

In order to determine the parameter $\kappa$, we apply the divergence theorem to Eq.~(\ref{eq:uequation})
\begin{equation}
\label{eq:divergence}
-\oint_{\partial \Omega} \nabla u \cdot \vec{n} \ dS = \int_\Omega \frac{1}{8} \bar{K}_{ij}\bar{K}^{ij} \psi^{-7} \ dV.
\end{equation}
Substituting $u=\kappa u_\text{g}$ (equivalently $h_\theta(x)=0$), we get the following equation for $\kappa$
\begin{equation}
\label{eq:divergence2}
-\kappa \oint_{\partial \Omega} \nabla u_{\text{g}} \cdot \vec{n}\ dS=
 \int_\Omega \frac{1}{8} \bar{K}_{ij}\bar{K}^{ij} \left( \psi_{\text{sing}} + \kappa u_{\text{g}} \right)^{-7} \ dV.
\end{equation}
Since $\kappa$ is used only to adjust the structure of the solution of $h_\theta(x)$, we do not need to find accurate solution of $\kappa$ to the above equation. We employ the quasi-Monte Carlo method \cite{MOROKOFF1995218} with uniformly distributed sampling points generated by Sobol sequence \cite{SOBOL196786} to calculate these integrals. For the volume integral, we use $5\times10^6$ points in the interior of the spherical domain. For the boundary integral, we use $5\times10^4$ points on the surface of the sphere.

\begin{figure}
    \centering
    \includegraphics[width=1.0\linewidth]{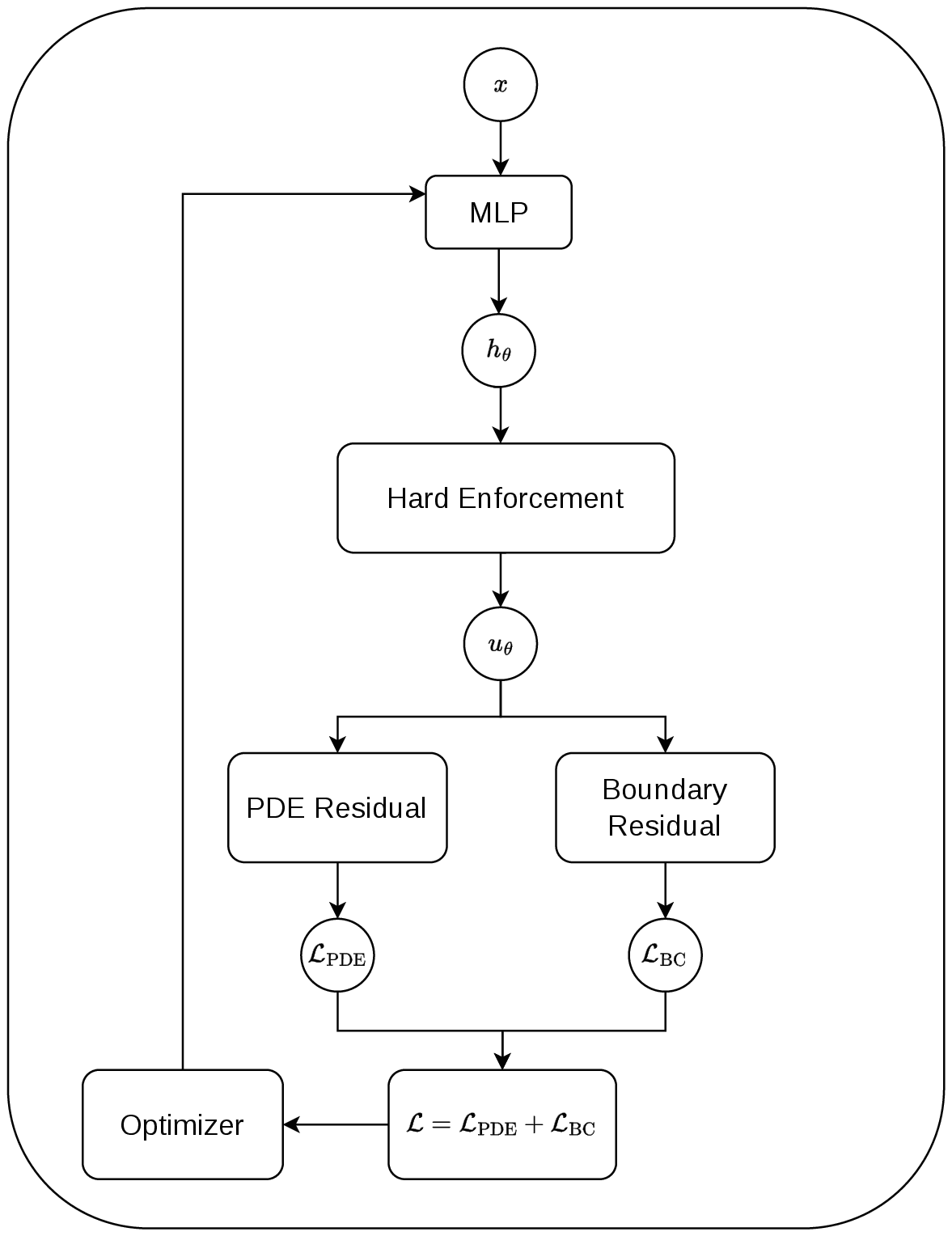}
    \caption{Schematic overview of the designed Physics-Informed Neural Network (PINN) framework in the current work for solving the Hamiltonian constraint equation. Input coordinates are processed through a multi-layer perceptron (MLP). The hard enforcement module imposes the analytical structural guidance (\ref{eq:ansatz}) onto the neural network output. The network is trained by minimizing the loss function (\ref{eq:totalloss}).}
    \label{fig:architecture}
\end{figure}

\begin{figure*}[!htbp]
    \centering
    \includegraphics[width=1.0\textwidth]{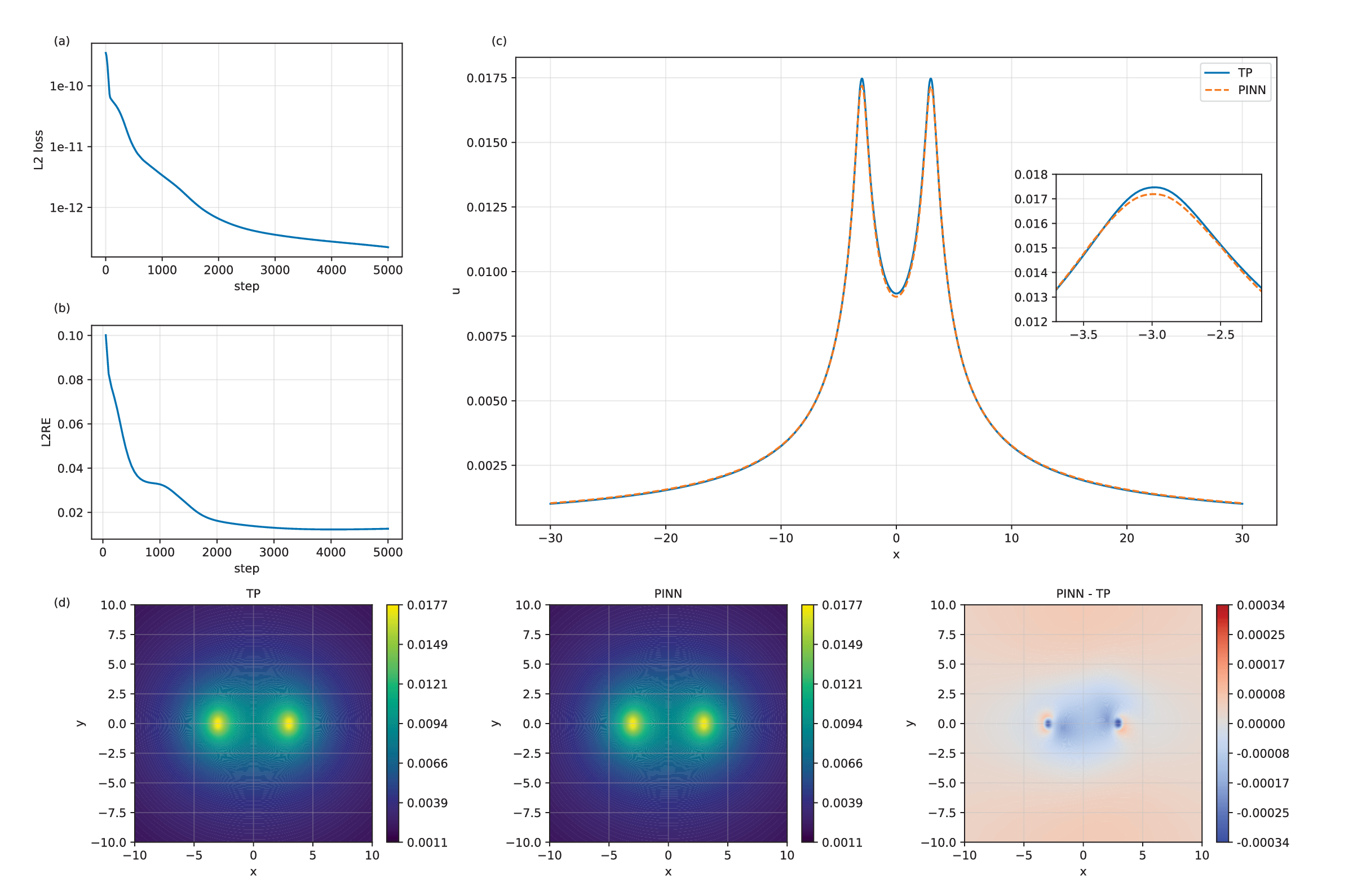}
    \caption{Behavior of our designed scheme about the solution of binary black hole with equal-mass holes $m_+=m_-=0.5$, placed on the x-axis at $x=\pm 3$, and with linear momenta $P^y_+=-P^y_-=0.2$. Panels (a) and (b) display the evolution of the $L_2$ loss and the L2RE over $5000$ training steps, respectively. Panel (c) compares the PINN solution against the \texttt{TwoPunctures} reference solution along the x-axis, with the inset highlighting the details at the $m_-$ puncture. Panel (d) compares the solutions on the $z\approx 0$ equatorial plane, showing the TP reference (left), the PINN prediction (middle), and the difference between the PINN solution and the TP reference solution (right).}
    \label{fig:ablationdefault}
\end{figure*}
\subsubsection{Training method}

\begin{figure*}[!htbp]
    \centering
    \includegraphics[width=1.0\textwidth]{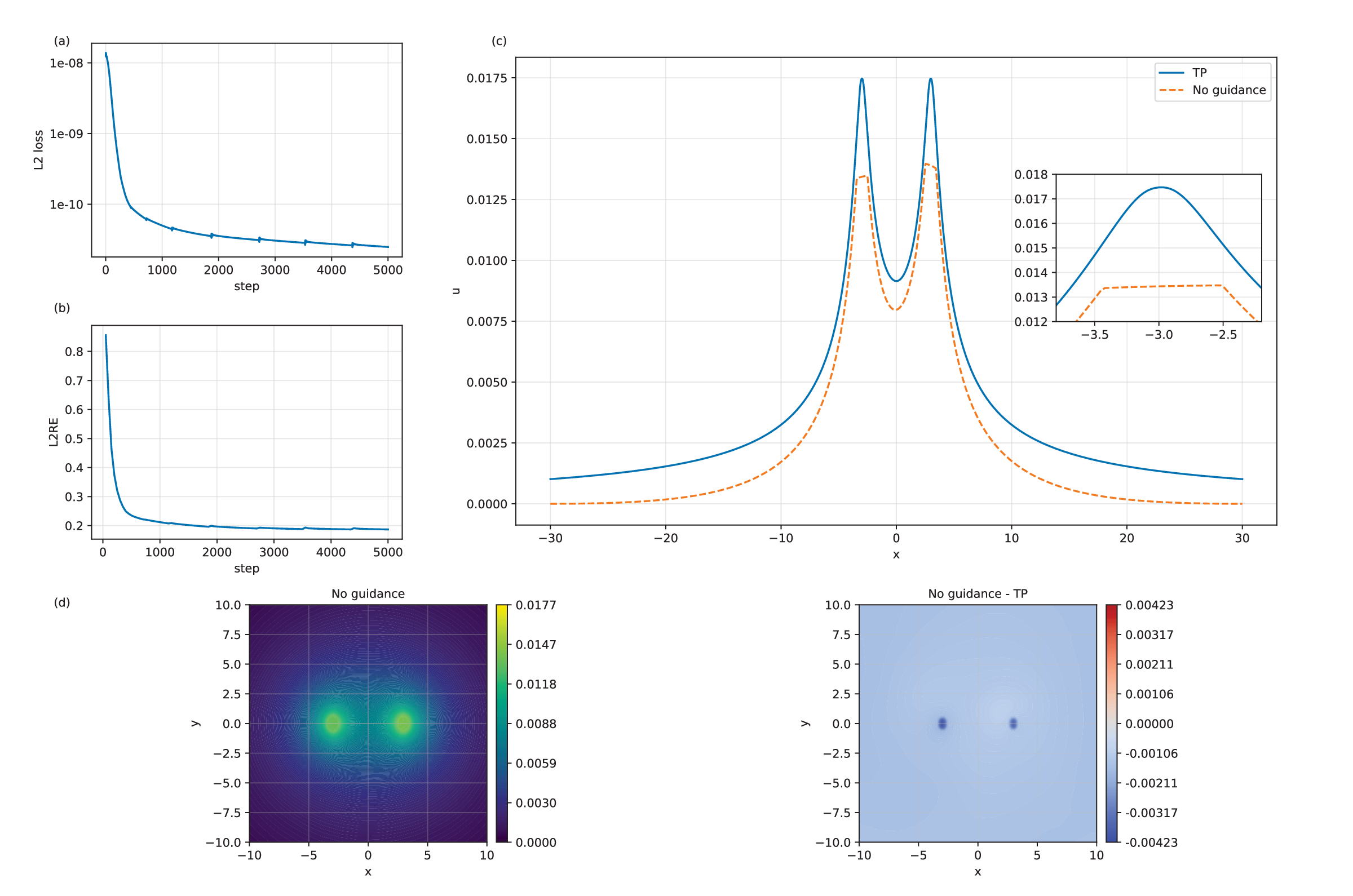}
    \caption{Similar to Fig.~\ref{fig:ablationdefault} but for the scheme without the analytical guidance. The TP reference solution is exactly the same as that shown in Fig.~\ref{fig:ablationdefault}, so here panel (c) only shows two subplots.}
    \label{fig:ablationnoguidance}
\end{figure*}
After determining the value of $\kappa$, we proceed to train $h_\theta(x)$ with the neural network. The network parameters $\theta$ are optimized by minimizing the loss function
\begin{equation}
\label{eq:totalloss}
\mathcal{L}=\mathcal{L}_{\text{PDE}}+\mathcal{L}_{\text{BC}}.
\end{equation}
For the PDE residual, we adopt the following form
\begin{equation}
\label{eq:pdeloss}
\mathcal{L}_{\text{PDE}}=w_2L_2+w_\infty\text{soft-}L_\infty.
\end{equation}
Corresponding to the residual of the equation $\mathcal{R}_i$ at each point, $L_2$ is defined as the mean squared error (MSE)
\begin{equation}
\label{eq:l2loss}
    L_2=\frac{1}{N_\Omega} \sum_{i=1}^{N_\Omega}{\mathcal{R}}_i^{2},
\end{equation}
and the $\text{soft-}L_\infty$ denotes the log-sum-exp (LSE) form of $L_\infty$ loss
\begin{equation}
\label{eq:softlinf}
\text{soft-}L_\infty=\frac{1}{\beta}\left(\ln\sum_{i=1}^{N_\Omega}e^{\beta |\mathcal R_i|}-\ln N_\Omega\right),
\end{equation}
where $\beta$ is used to control the sharpness of the approximation. Larger $\beta$ makes $\text{soft-}L_\infty$ closer to the true $L_\infty$ norm by emphasizing the maximum residual. This smooth formulation both constrains large residuals and avoids the non-differentiability of the exact $L_\infty$ loss. $w_2$ and $w_\infty$ are the weights of $L_2$ and $\text{soft-}L_\infty$ respectively.

For boundary residuals we impose a Robin boundary condition (\ref{robin}). Given the boundary residual $\mathcal{R}_{B,i}$, the corresponding loss is given by
\begin{equation}
\label{eq:robinloss}
\mathcal{L}_\text{BC}=w_{\text{rob}}\cdot
\frac{1}{N_{\partial\Omega}}\sum_{i=1}^{N_{\partial\Omega}}\mathcal{R}_{B,i}^2,
\end{equation}
where $w_{\text{rob}}$ is used to control the weight of the boundary loss.

Parameters $w_2$, $w_\infty$, $\beta$ and $w_{\text{rob}}$ are pre-fixed before training. Before training, we uniform-randomly choose $N_\Omega = 20000$ inner points and $N_{\partial\Omega} = 8000$ boundary points in the current work.

\subsubsection{Loss balancing strategy}
The total loss function contains three components including $L_k=\left\{L_2,\text{soft-}L_\infty,\mathcal{L}_\text{BC}\right\}$, which may differ by several orders of magnitude. When combined directly, the largest magnitude term may dominate the optimization and hinder the reduction of other constraints. To mitigate this scale imbalance, we dynamically rescale each individual loss term using an exponential moving average (EMA) method.

For each loss term $L_k$ at training step $t_n$, the EMA of each loss $\bar{L}_k(t_n)$ is given by
\begin{equation}
\label{eq:ema}
\bar{L}_k(t_n) = \alpha \bar{L}_k(t_{n-1}) + (1-\alpha)L_k(t_n)
\end{equation}
where $\alpha\in[0,1)$ is a tunable hyperparameter. The rescaled loss $\tilde{L}_k$ is then calculated by
\begin{align}
\label{eq:lossbalancing}
\tilde{L}_k(t_n) &= \frac{L_k(t_n)}{\bar{L}_k(t_n)}.
\end{align}
This strategy rescales each loss term so that different loss components remain at comparable numerical scales during training. As a result, no single large-magnitude term dominates the optimization solely, which improves the stability and effectiveness of the multi-loss training process.

\subsubsection{Strategy summary}

We summarize the architecture of our PINN method for Hamiltonian constraint equations in Fig.~\ref{fig:architecture}. The network consists of $3$ hidden layers. Each layer has 64 neurons. Every neuron is equipped with a sigmoid linear unit (SiLU) activation function \cite{hendrycks2016gaussian, elfwing2018sigmoid}, $\mathrm{SiLU}(x)=x/(1+e^{-x})$. The network is trained for $5000$ steps using the adaptive moment estimation (Adam) optimizer \cite{kingma2014adam} with a learning rate of $3\times 10^{-4}$.

As a typical example, we consider a binary black hole system with two equal-mass holes with $m_+=m_-=0.5$, placed on x-axis at $x=\pm 3$, and linear momenta $P^y_+=-P^y_-=0.2$. To solve this system we set the hyperparameters as $c=0.2$, $w_2=1.0$, $w_\infty=0$, and $w_{\text{rob}}=1.0$.

In order to quantitatively analyze the error, we evaluate the relative $L_2$ error (L2RE) of the PINN predicted solution on the spectral grid nodes of \texttt{TwoPunctures}, and make comparison with the reference solution. The L2RE is defined as
\begin{equation}
\label{eq:l2re}
\text{L2RE} = \sqrt{\frac{\sum_{i=1}^{N_{\text{grid}}} |u_{\text{PINN}}(\mathbf{x}_i) - u_{\text{TP}}(\mathbf{x}_i)|^2}{\sum_{i=1}^{N_{\text{grid}}} |u_{\text{TP}}(\mathbf{x}_i)|^2}},
\end{equation}
where $\text{TP}$ means the solution got by \texttt{TwoPunctures}.

\begin{table}[!tbp]
    \centering
    \renewcommand{\arraystretch}{1.4}
    \setlength\tabcolsep{15 pt}
    \caption{Quantitative comparison of the L2RE defined in (\ref{eq:l2re}) between our designed scheme and the schemes without key techniques. The evaluations are performed on a binary black hole system with equal masses $m_+=m_-=0.5$, placed on the x-axis at $x=\pm 3$, and with linear momenta $P^y_+=-P^y_-=0.2$. For each strategy we train the network independently $5$ times with different initial random seeds. The optimal performance achieved by the designed scheme is marked in bold. The number before $\pm$ corresponds to the average value of the 5 random runs. The number after $\pm$ denotes the standard deviation of these 5 runs.}
    \begin{tabular}{lc}
        \hline
        \textbf{Configuration} & \textbf{L2RE} \\
        \hline
        Designed scheme & $\mathbf{0.0172 \pm 0.0040}$ \\
        without guidance  & $0.2417 \pm 0.0397$ \\
        guidance form (\ref{eq:add})      & $0.0363 \pm 0.0087$ \\
        alternative window (\ref{eq:smoothwindow})  & $0.0866 \pm 0.0079$ \\
        $\kappa=1$      & $0.5357 \pm 0.0470$ \\
        unbalanced    & $0.1039 \pm 0.0097$ \\
        \hline
    \end{tabular}
    \label{tab:ablation}
\end{table}

\begin{figure*}[!htbp]
    \centering
    \includegraphics[width=1.0\textwidth]{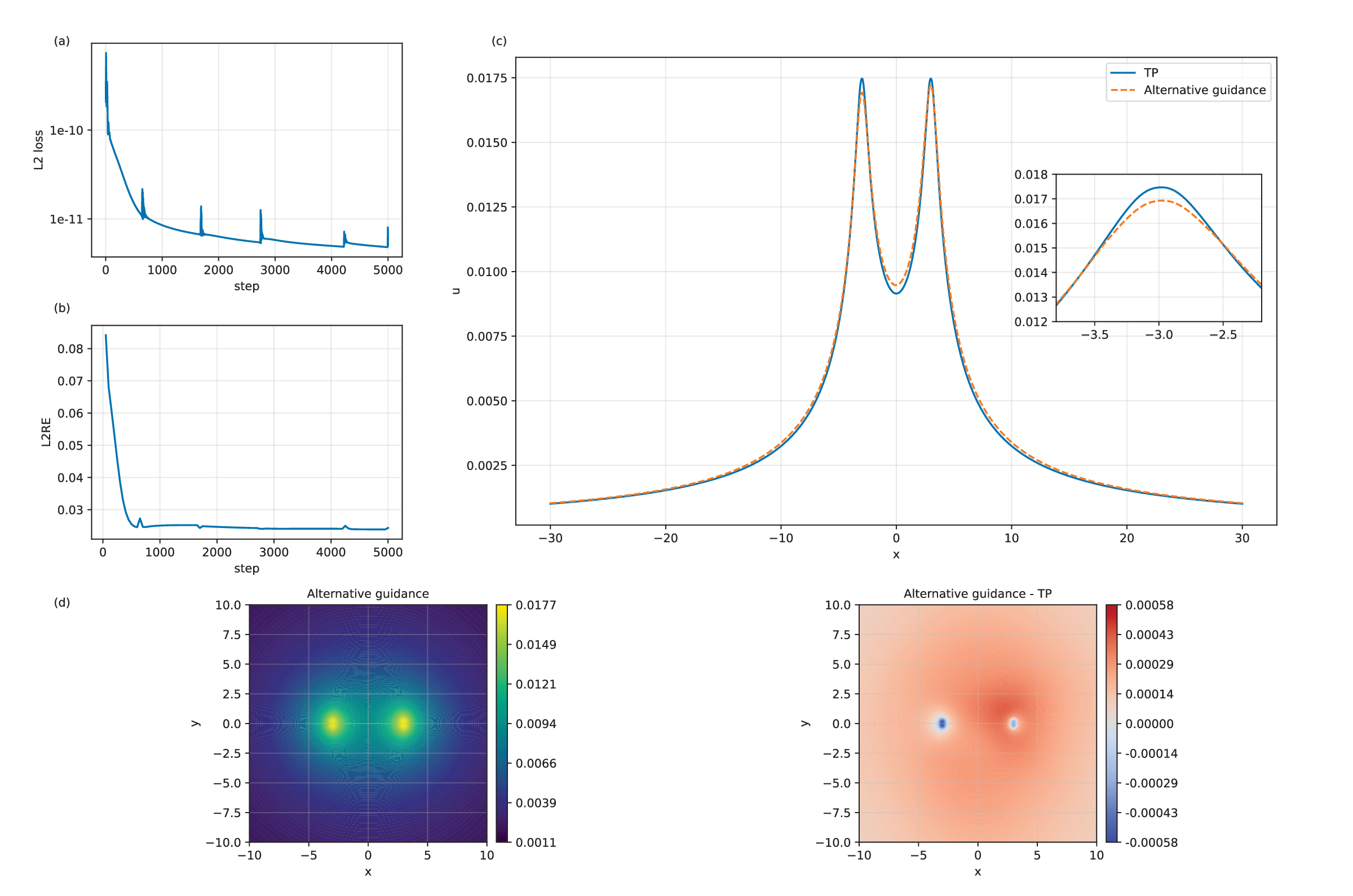}
    \caption{Same to Fig.~\ref{fig:ablationnoguidance} but for the scheme with an alternative analytical guidance form (\ref{eq:add}).}
    \label{fig:ablationadd}
\end{figure*}

In order to check the robustness of our strategy we train the network independently $5$ times with different initial random seeds. The resulted L2RE is presented in the first row of Table~\ref{tab:ablation}. The number before $\pm$ corresponds to the average value of the 5 random runs. The number after $\pm$ is the standard deviation of these 5 runs. We show the solution behavior of one of these 5 runs in Fig.~\ref{fig:ablationdefault}. As reported in the table, our designed scheme can achieve about 1\% accuracy. In Fig.~\ref{fig:ablationdefault}a, the $L_2$ loss decreases monotonically throughout the training process, indicating the stable optimization behavior. Meanwhile, Fig.~\ref{fig:ablationdefault}b shows that the L2RE drops rapidly during the early stage (within the first few hundred steps) and then continues to decrease more gradually, eventually reaching a stable plateau around the $10^{-2}$ level without noticeable oscillations. Fig.~\ref{fig:ablationdefault}c compares the PINN solution with the \texttt{TwoPunctures} reference solution along the x-axis. The two curves overlap closely over the entire domain including the near-field region around the punctures. The inset subplot zooms in the neighborhood near the puncture point. Fig.~\ref{fig:ablationdefault}d presents the comparison on the $z=0$ plane. The PINN solution reproduces the overall \texttt{TwoPunctures} reference solution. The main deviation is localized primarily around the strong-field regions near the punctures.

\begin{figure*}[!htbp]
    \centering
    \includegraphics[width=1.0\textwidth]{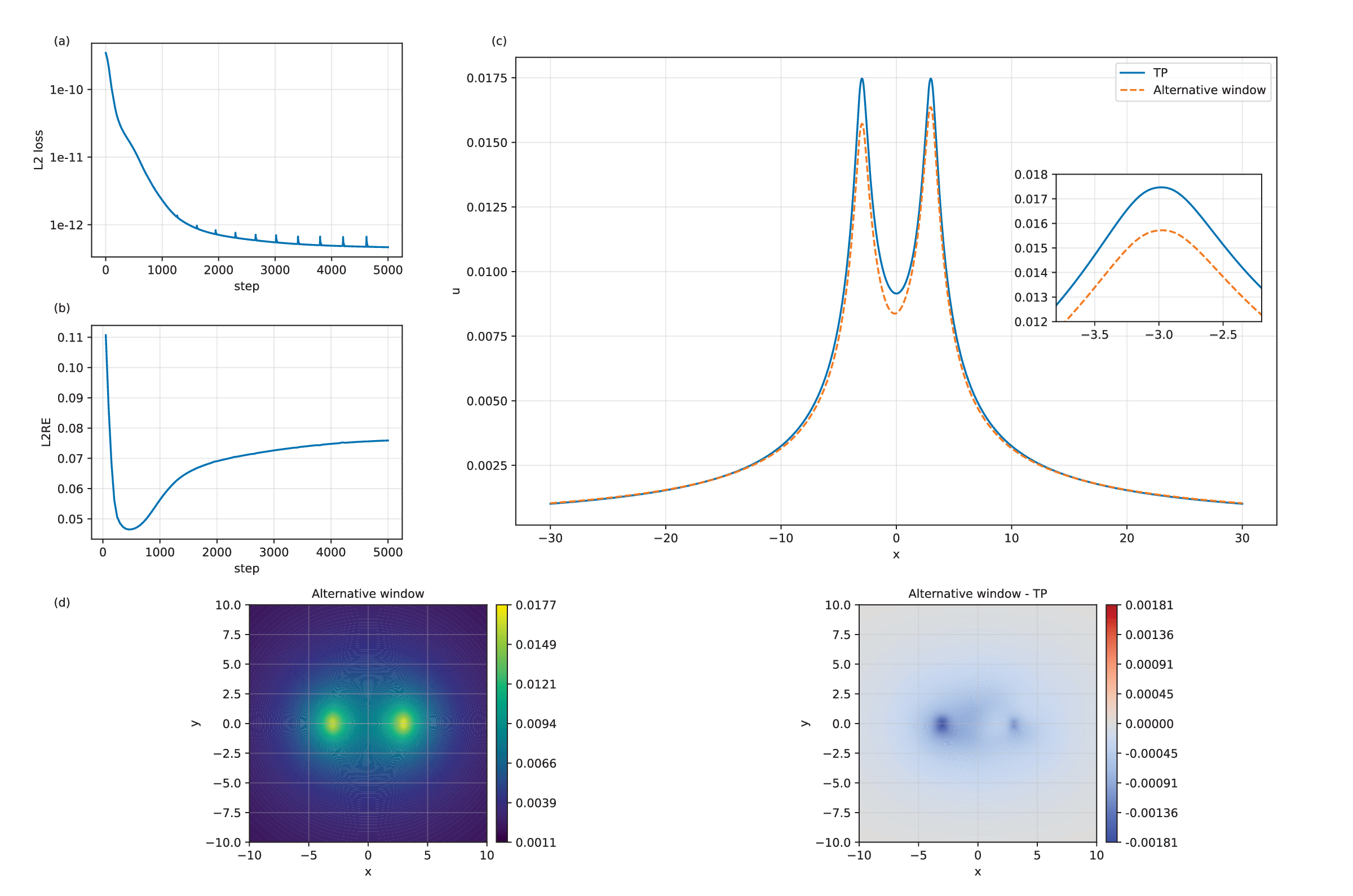}
    \caption{Same to Fig.~\ref{fig:ablationnoguidance} but for the scheme with an alternative general smooth window.}
    \label{fig:ablationsmoothwindow}
\end{figure*}

\begin{figure*}[!htbp]
    \centering
    \includegraphics[width=1.0\textwidth]{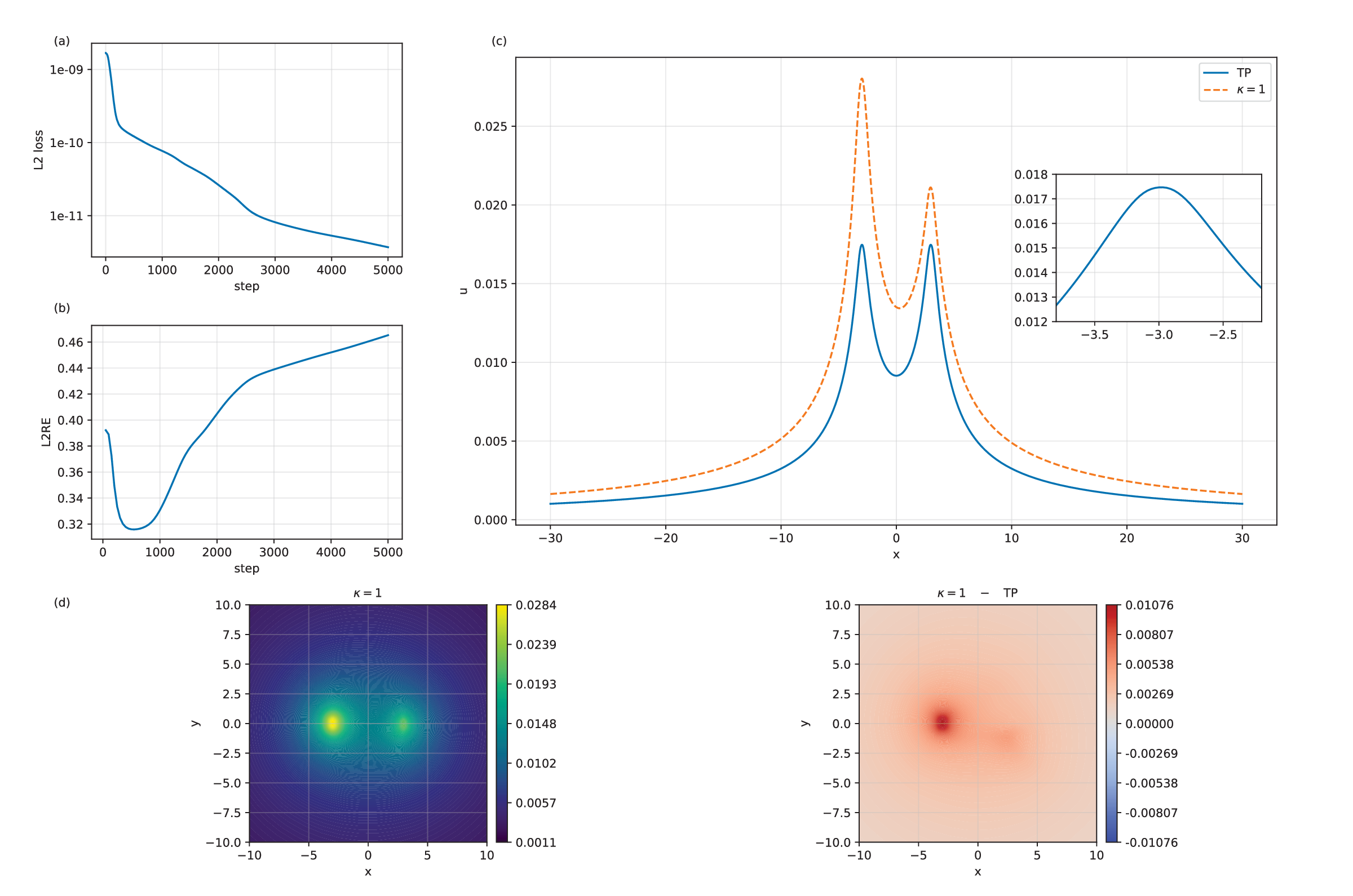}
    \caption{Same to Fig.~\ref{fig:ablationnoguidance} but for the scheme with $\kappa = 1$.}
    \label{fig:ablationnostage1}
\end{figure*}

\section{The effect of the techniques in our designed scheme}
\label{s3}
Our designed scheme includes two important blocks: the analytical guidance (\ref{eq:ansatz}) and the loss balancing strategy. In the following we show the important effect of these two blocks.

\subsection{Effect of the analytical guidance}
\label{s3a}
By removing the guiding solution, the physical ansatz reduces to
\begin{equation}
\label{eq:noguidance}
u_{\theta,\text{ng}}(x) = c W(x)\tanh(h_\theta(x)).
\end{equation}
As shown in the second row of Table \ref{tab:ablation}, the removal of analytical guidance leads to dramatic performance degradation, yielding an average L2RE of $0.2417 \pm 0.0397$, which is more than an order of magnitude higher than our designed scheme. Again we randomly run 5 times and the average value and the standard deviation are presented here.

One of the 5 runs' results is shown in Fig.~\ref{fig:ablationnoguidance} which is similar to Fig.~\ref{fig:ablationdefault}. The evolutions of the $L_2$ loss and the L2RE are depicted in Fig.~\ref{fig:ablationnoguidance}a and Fig.~\ref{fig:ablationnoguidance}b, respectively. Although the $L_2$ loss decreases steadily during training, the L2RE only drops rapidly in the first few hundred steps and then gradually approaches a relatively high plateau. This indicates that, without the analytical guidance, minimizing the residual on collocation points does not result in an accurate global solution on the \texttt{TwoPunctures} spectral grid. As shown in Fig.~\ref{fig:ablationnoguidance}c, the configuration without analytical guidance captures the overall trend but noticeably underestimates the solution amplitude in the strong-field region near the punctures. The inset subplot further highlights the mismatch around the puncture neighborhood. Fig.~\ref{fig:ablationnoguidance}d shows that, the large-scale spatial pattern can be reproduced but the solution is generally smaller than the true solution. The largest differences concentrated near the two puncture points.

\begin{figure*}[!htbp]
    \centering
    \includegraphics[width=1.0\textwidth]{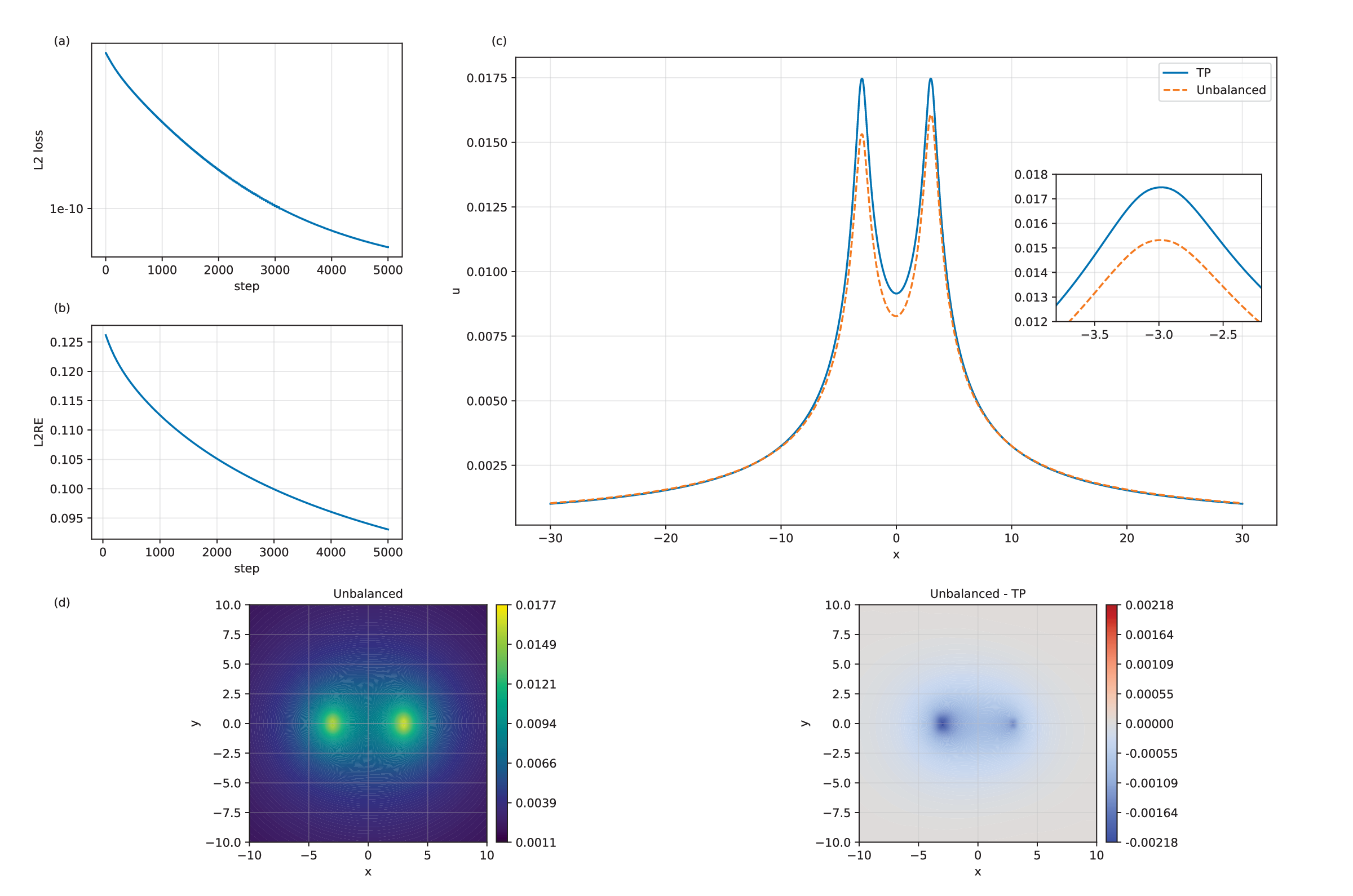}
    \caption{Same to Fig.~\ref{fig:ablationnoguidance} but for the scheme without the loss balancing strategy.}
    \label{fig:ablationunbalanced}
\end{figure*}
\subsubsection{Impact of the correction form}
\label{s3c}
We have seen the important effect of the analytical guidance to the numerical solution above. In addition, we find that the guidance form is also important. To show this effect, we replace the ansatz (\ref{eq:ansatz}) with an alternative form
\begin{equation}
\label{eq:add}
u_{\theta,\text{add}}(x) = \kappa\,u_{\text{g}}(x)+cW(x) h_\theta(x).
\end{equation}

As indicated in the third row of Table~\ref{tab:ablation}, the alternative analytical guidance form yields an average L2RE of $0.0363 \pm 0.0087$. Once again 5 random runs are used. While it significantly outperforms the scheme without a guiding solution (\ref{eq:noguidance}), its error is significantly higher than that of our designed scheme.

Similar to Fig.~\ref{fig:ablationnoguidance}, we show one example of the 5 runs in Fig.~\ref{fig:ablationadd}. The $L_2$ loss decreases rapidly at the beginning and continues to decay throughout training, with several intermittent spikes during the optimization process. The L2RE drops quickly within the first few hundred steps and then gradually converges to a stable plateau around the $10^{-2}$ level. Compared with our designed scheme, the additive analytical guidance form exhibits a higher final error and a slower late-stage improvement, indicating reduced effectiveness of error correction in the strong-field region.

The comparisons along the x-axis and on the $z=0$ plane are presented in Fig.~\ref{fig:ablationadd}c and in Fig.~\ref{fig:ablationadd}d respectively. As shown in Fig.~\ref{fig:ablationadd}c, although the alternative analytical guidance form works over the entire domain, it works not as good as our designed scheme in the strong-field regions near the punctures. The inset further highlights the solution difference between the alternative analytical guidance form and our designed form in the neighborhood of the $m_-$ puncture. Fig.~\ref{fig:ablationadd}d shows that the difference field remains small and is mainly concentrated around the two punctures and in the region between them.

\subsubsection{Impact of the Window Function}
\label{s3d}

In our designed scheme, we use a window function based on the magnitude of the guiding solution. To evaluate the necessity of this approach, we compare it against a general smooth window $W^\prime(x)$, which is constructed through the superposition of individual smooth window functions $S_n$ for each puncture
\begin{equation}
\label{eq:smoothwindow}
S_n(r_n) = \frac{1}{2}\left[1 - \tanh\left(\frac{r_n - r_c}{\delta}\right)\right],
\end{equation}
where $r_n$ represents the spatial distance to the $n$-th puncture, $r_c$ is the radius of the window function, and $\delta$ controls the width of the smooth transition region.

Replacing $W(x)$ with $W^\prime(x)$ results in an average L2RE of $0.0866 \pm 0.0079$, as reported in the fourth row of Table~\ref{tab:ablation}, which is also a noticeable performance degradation compared to our designed scheme.

Again, one example evolution of the 5 runs for the $L_2$ loss and the L2RE are presented in Fig.~\ref{fig:ablationsmoothwindow}a and Fig.~\ref{fig:ablationsmoothwindow}b. While the $L_2$ loss decreases steadily throughout training, the L2RE exhibits a non-monotonic behavior. The L2RE drops rapidly at the beginning and reaches its minimum within the first $500$ steps, but then gradually increases and saturates at a noticeably high level. This indicates that under a general smooth window setting, further reduction of the collocation residual does not translate into improved global accuracy on the \texttt{TwoPunctures} spectral grid, and the optimization becomes progressively biased away from the true solution.

The comparisons along the x-axis and on the $z=0$ plane are shown in Fig.~\ref{fig:ablationsmoothwindow}c and Fig.~\ref{fig:ablationsmoothwindow}d, respectively. In Fig.~\ref{fig:ablationsmoothwindow}c, the solution obtained with a general smooth window setting follows the overall solution profile of our designed scheme, but it systematically underestimates the amplitudes near both punctures. The inset highlights a clear discrepancy in the neighborhood of the $m_-$ puncture. Consistently, Fig.~\ref{fig:ablationsmoothwindow}d shows that the difference is mainly concentrated around the two strong-field regions near the punctures, whereas the discrepancy in the far field remains comparatively small.

\subsubsection{Impact of the factor $\kappa$}

In order to investigate the impact of the factor $\kappa$ in our designed scheme, we simply set $\kappa = 1$ and train $h_\theta(x)$ directly. As reported in the fifth row of Table~\ref{tab:ablation}, $\kappa=1$ leads to an average L2RE of $0.5357 \pm 0.0470$, which is significantly higher than that of our designed scheme.

Again, one example evolution of the 5 runs for $L_2$ loss function and L2RE are presented in Fig.~\ref{fig:ablationnostage1}a and Fig.~\ref{fig:ablationnostage1}b. Both the $L_2$ loss function and the L2RE are significantly larger than that of our designed scheme. Fig.~\ref{fig:ablationnostage1}b also shows that the L2RE drops at the beginning and increases in the later process. At the end of training, the L2RE is even higher than its initial value. This behavior shows that, the network with $\kappa=1$ cannot obtain the correct boundary value, which severely degrades the effectiveness of the subsequent training.

The comparisons along the x-axis and on the $z=0$ plane are presented in Fig.~\ref{fig:ablationnostage1}c and in Fig.~\ref{fig:ablationnostage1}d respectively. The results shows that the numerical solution exhibits large deviations throughout the computational domain, demonstrating that the training is unsuccessful in this setting.

\begin{figure*}[!htbp]
    \centering
    \includegraphics[width=1.0\textwidth]{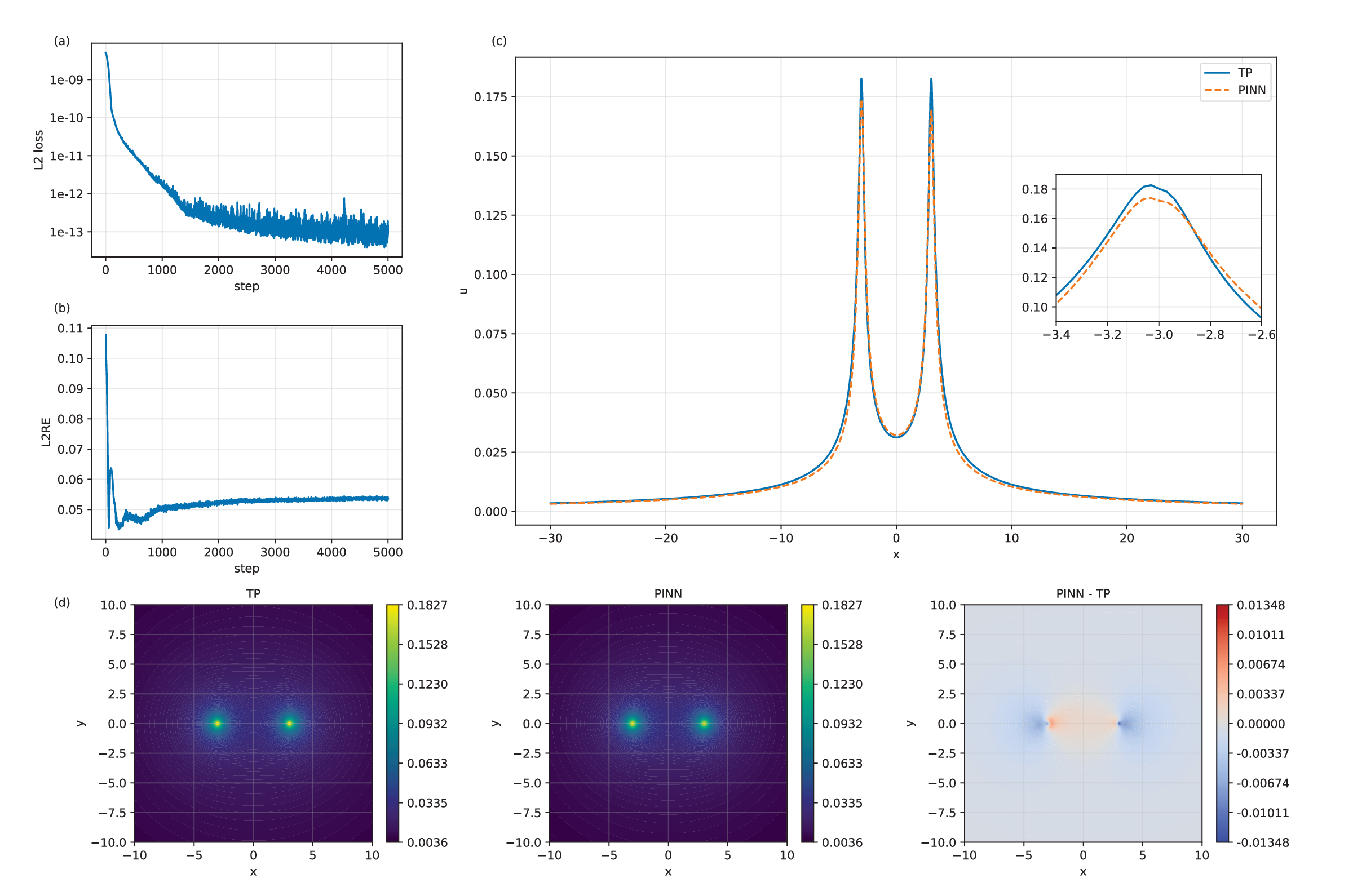}
    \caption{Similar to Fig.~\ref{fig:ablationdefault} but for an equal-mass spinning binary black hole system ($m_+=m_-=0.5$, $P^y_+=-P^y_-=0.2$, $S^z_+=S^z_-=0.2$ at $x=\pm 3$).}
    \label{fig:resultsequalspinning}
\end{figure*}
\subsection{Effect of the loss balancing strategy}
\label{s3f}
In this subsection, we investigate the impact of the loss balancing strategy. To do so we optimize the composite loss with fixed weights throughout training. As reported in the last row of Table~\ref{tab:ablation}, removing the balancing strategy leads to a clear performance degradation, yielding an average L2RE of $0.1039 \pm 0.0097$, which is significantly higher than that of our designed scheme.

As above, one example evolution of the 5 runs for $L_2$ loss function and L2RE are presented in Fig.~\ref{fig:ablationunbalanced}a and \ref{fig:ablationunbalanced}b. Without the balancing strategy, the L2RE decreases noticeably slower than that of our designed scheme, indicating a reduced optimization efficiency when the loss components remain unnormalized.

The comparisons along the x-axis and on the $z=0$ plane are presented in Fig.~\ref{fig:ablationunbalanced}c and in Fig.~\ref{fig:ablationunbalanced}d respectively. As shown in Fig.~\ref{fig:ablationunbalanced}c, the unbalanced configuration reproduces the overall profile of the default solution, but it systematically underestimates the amplitudes near the punctures. The inset highlights a more noticeable discrepancy in the neighborhood of the $m_-$ puncture. Consistently, Fig.~\ref{fig:ablationunbalanced}d shows that the difference is mainly concentrated around the two strong-field regions near the punctures, with a larger deviation around the $m_-$ puncture, while the discrepancy in the far field remains comparatively small.

\begin{figure*}[!htbp]
    \centering
    \includegraphics[width=1.0\textwidth]{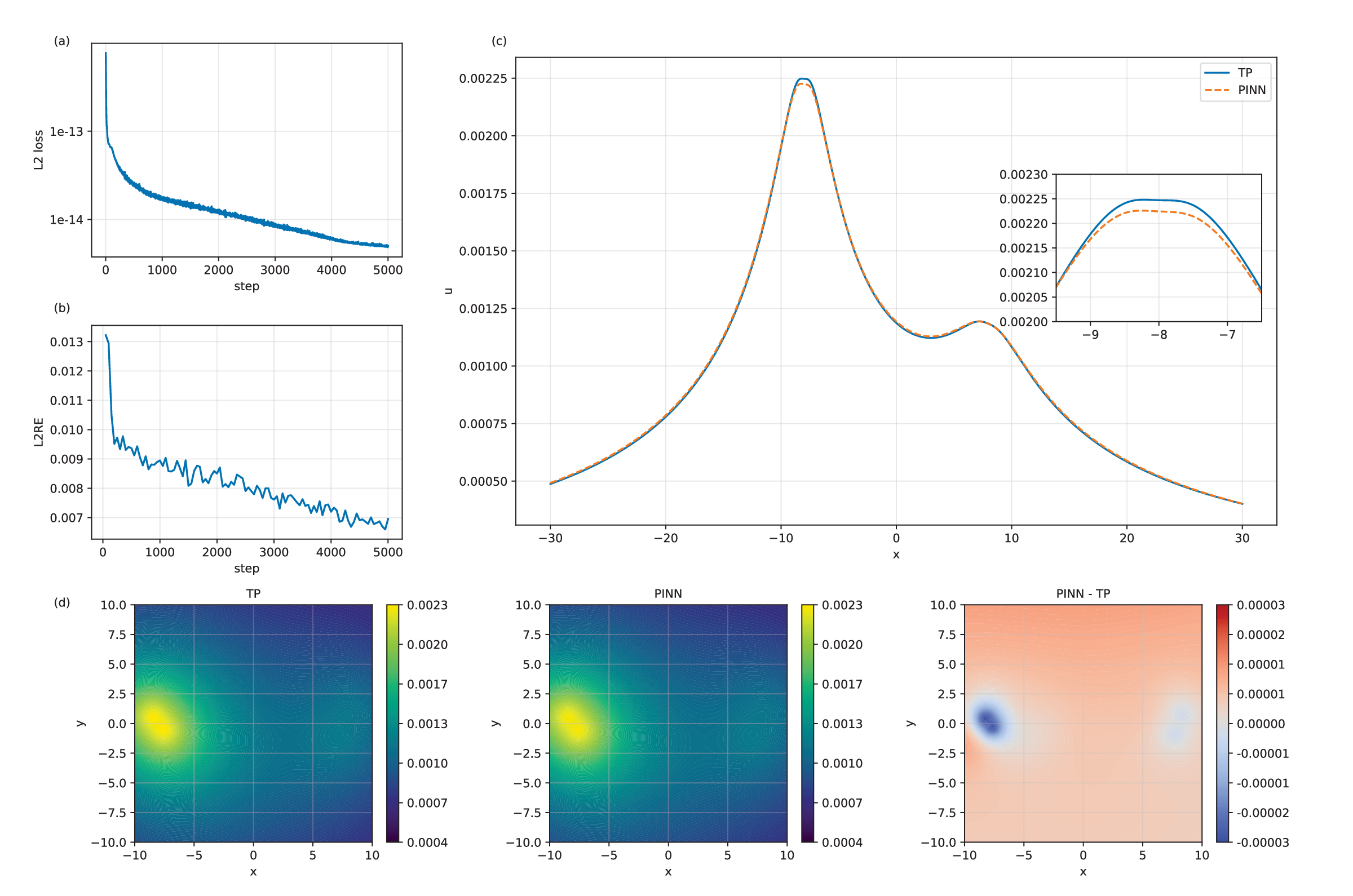}
    \caption{Similar to Fig.~\ref{fig:ablationdefault} but for an unequal-mass non-spinning binary black hole system ($m_+=3, m_-=2$, $P^x_+=0.1$, $P^y_+=0.25$, $P^x_-=0.2$, $P^y_-=-0.25$ at $x=\pm 8$).}
    \label{fig:resultsunequalnonspinning}
\end{figure*}

\section{Solutions for typical binary black hole initial data with our designed scheme}
\label{s4}
In this section, we apply our designed scheme to the initial data problem for typical binary black hole configurations. Specifically, an equal-mass spinning binary, an unequal-mass non-spinning binary and an unequal-mass spinning binary are considered.
\begin{figure*}[!htbp]
    \centering
    \includegraphics[width=1.0\textwidth]{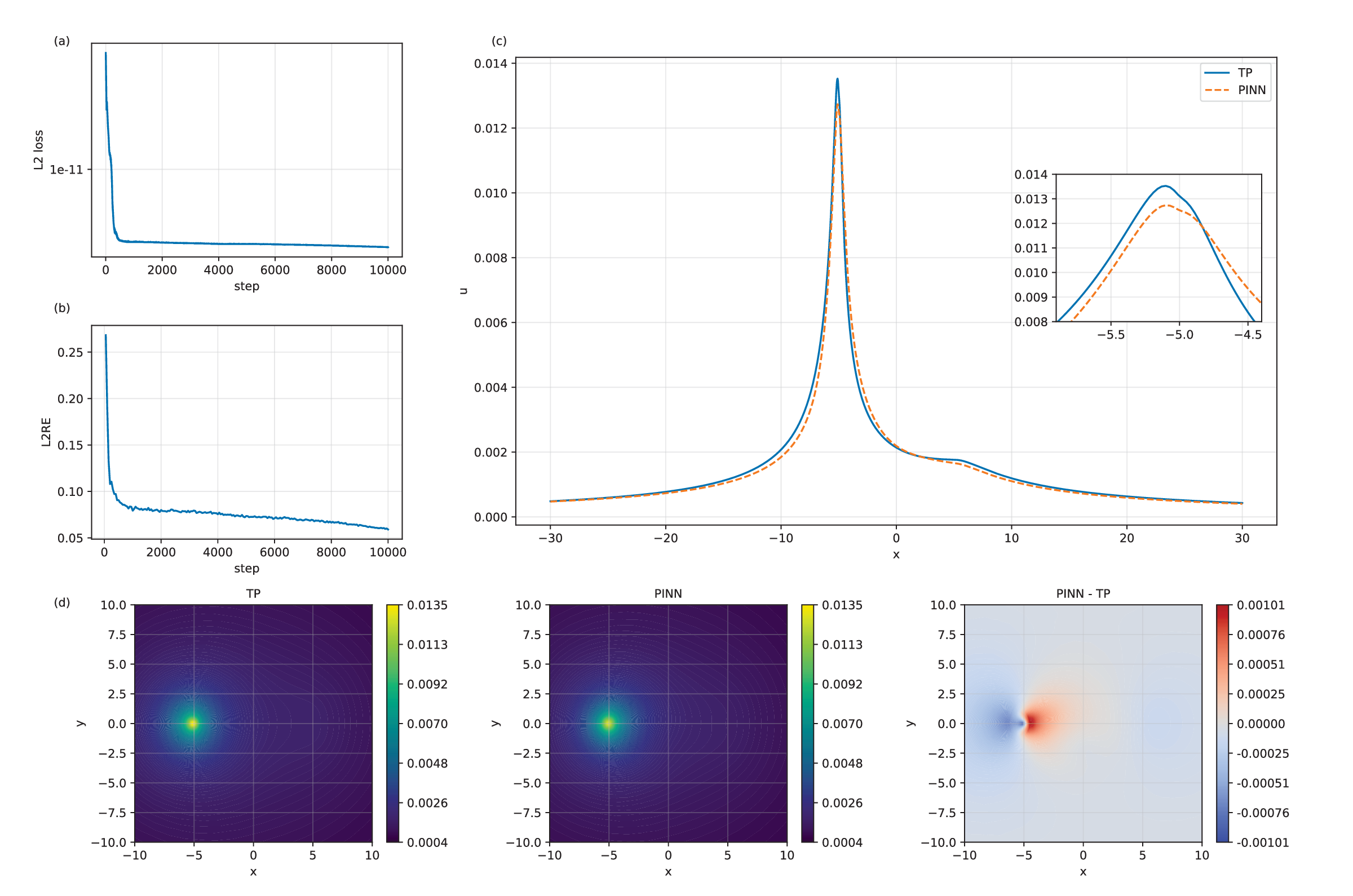}
    \caption{Similar to Fig.~\ref{fig:ablationdefault} but for an unequal-mass, spinning binary black hole system ($m_+=3, m_-=1$, $ P^y_+=0.3$, $P^y_-=-0.2$, $S^z_+=0.1$ and $S^z_-=0.2$ at $x=\pm 5$).}
    \label{fig:resultsunequalspinning}
\end{figure*}
\subsection{Equal-mass spinning binary}
\label{s4a}
In this section, we consider an equal-mass spinning binary system with $m_+=m_-=0.5$, placed on x-axis at $x=\pm 3$, with linear momenta $P^y_+=-P^y_-=0.2$, and spins $S^z_+=S^z_-=0.2$. The learning rate is set to $5\times 10^{-4}$ for $5000$ training steps. Other hyperparameters are set to $c=1.0$, $w_2=1.0$, $w_\infty=0.5$, $\beta=2.5$, and $w_{\text{rob}}=1.0$.

\begin{table}[!htb]
    \centering
    \renewcommand{\arraystretch}{1.4}
    \setlength\tabcolsep{10 pt}
    \caption{Quantitative behavior of the L2RE for three typical binary black hole configurations including an equal-mass spinning system ($m_+=m_-=0.5$, $P^y_+=-P^y_-=0.2$, $S^z_+=S^z_-=0.2$ at $x=\pm 3$), an unequal-mass non-spinning system ($m_+=3, m_-=2$, $P^x_+=0.1, P^y_+=0.25$, $P^x_-=0.2, P^y_-=-0.25$ at $x=\pm 8$) and an unequal-mass spinning system ($m_+=3, m_-=1$, $P^y_+=0.3$, $P^y_-=-0.2$, $S^z_+=0.1$, $S^z_-=0.2$ at $x=\pm 5$). We take $5$ independent runs for each case. As shown in Table~\ref{tab:ablation}, the average value and the standard deviation are presented before and after the $\pm$ sign.}
    \begin{tabular}{lc}
        \hline
        \textbf{Configuration} & \textbf{L2RE} \\
        \hline
        Equal-mass spinning    &   $0.0838 \pm 0.0319$ \\
        Unequal-mass non-spinning  &        $0.0077 \pm 0.0007$ \\
        Unequal-mass spinning      &        $0.0889 \pm 0.0250$ \\
        \hline
    \end{tabular}
    \label{tab:results}
\end{table}
As reported in Table~\ref{tab:results}, the network achieves an L2RE of $0.0838 \pm 0.0319$ for the equal-mass spinning case over $5$ independent runs. Compared with the non-spinning cases considered previously, the obtained error is noticeably larger, indicating that the presence of spin introduces additional difficulties for the proposed PINN framework. As last section, one example evolution of the 5 runs for the $L_2$ loss and the L2RE in this case are presented in Fig.~\ref{fig:resultsequalspinning}a and Fig.~\ref{fig:resultsequalspinning}b. As shown in Fig.~\ref{fig:resultsequalspinning}a, the $L_2$ loss decreases rapidly at the beginning and continues to decay overall, with small fluctuations at later training steps. Correspondingly, Fig.~\ref{fig:resultsequalspinning}b shows that the L2RE drops quickly at the beginning, followed by a mild increase and eventually stabilizes around $5\times10^{-2}$.

The comparisons along the x-axis and on the $z=0$ plane are presented in Fig.~\ref{fig:resultsequalspinning}c and Fig.~\ref{fig:resultsequalspinning}d, respectively. Fig.~\ref{fig:resultsequalspinning}c shows that the PINN solution captures the overall structure of the solution, while a visible deviation remains near the peak regions, as highlighted in the inset. In Fig.~\ref{fig:resultsequalspinning}d, the PINN solution reproduces the main spatial pattern of the reference solution, while the difference map indicates that the absolute error remains relatively small in the entire region.

\subsection{Unequal-mass non-spinning binary}
\label{s4b}
In this section, we consider an unequal-mass non-spinning binary system with $m_+=3, m_-=2$, $P^x_+=0.1, P^y_+=0.25$, $P^x_-=0.2, P^y_-=-0.25$ at $x=\pm 8$. The learning rate is set to $5\times 10^{-4}$ for $5000$ training steps. Other hyperparameters are set to $c=1.0$, $w_2=1.0$, $w_\infty=0.5$, $\beta=10$ and $w_{\text{rob}}=1.0$.

As reported in the second row of Table \ref{tab:results}, the result yields highly robust and accurate solutions across $5$ independent runs, achieving an extremely small average L2RE of $0.0077 \pm 0.0007$. As last section, one example evolution of the 5 runs for the $L_2$ loss function and the L2RE in this case are presented in Fig.~\ref{fig:resultsunequalnonspinning}a and \ref{fig:resultsunequalnonspinning}b. As shown in the figure, the $L_2$ loss exhibits a steady and continuous descent throughout the majority of the optimization process. Correspondingly, while the L2RE curve exhibits more fluctuations compared to the previous cases, the numerical value of the L2RE itself has already dropped to an exceptionally low level, and it consistently maintains a robust and stable downward trend across the entire training process.

The comparisons along the x-axis and on the $z=0$ plane are presented in Fig.~\ref{fig:resultsunequalnonspinning}c and in Fig.~\ref{fig:resultsunequalnonspinning}d, respectively. It can be seen that the PINN solution exhibits excellent agreement with the \texttt{TwoPunctures} reference solution. Specifically, the 1D profile confirms that the network perfectly captures the behavior of the solution, without any noticeable deviations. Furthermore, the 2D error distribution demonstrates that the maximal absolute differences are highly suppressed across the entire computational domain, with the magnitude of roughly $2 \times 10^{-5}$.

\subsection{Unequal-mass spinning binary}
\label{s4c}
In this subsection, we consider an unequal-mass spinning binary system with $m_+=3, m_-=1$, $ P^y_+=0.3$, $P^y_-=-0.2$, $S^z_+=0.1$ and $S^z_-=0.2$ at $x=\pm 5$. For this configuration, the learning rate is set to $5\times 10^{-4}$ for $10000$ training steps. Other hyperparameters are set to $c=1.0$, $w_2=1.0$, $w_\infty=1.0$, $\beta=10$ and $w_{\text{rob}}=1.0$.

As reported in the last row in the Table~\ref{tab:results}, the network yields relatively accurate predictions across $5$ independent runs, achieving an average L2RE of $0.0889 \pm 0.0250$. The example evolutions of the $L_2$ loss and the L2RE in this case are presented in Fig.~\ref{fig:resultsunequalspinning}a and Fig.~\ref{fig:resultsunequalspinning}b. As shown in Fig.~\ref{fig:resultsunequalspinning}a, the $L_2$ loss decreases steadily over the training process, indicating stable optimization. Correspondingly, Fig.~\ref{fig:resultsunequalspinning}b shows that the L2RE drops rapidly during the early stage and then decreases more gradually with mild fluctuations, eventually plateauing at a relatively higher level than the non-spinning case, reflecting the increased difficulty of this configuration.

The comparisons along the x-axis and on the $z=0$ plane are presented in Fig.~\ref{fig:resultsunequalspinning}c and Fig.~\ref{fig:resultsunequalspinning}d, respectively. Fig.~\ref{fig:resultsunequalspinning}c shows that the PINN solution captures the overall shape of the \texttt{TwoPunctures} reference solution, especially in the far-field region, but noticeable discrepancies remain near the puncture neighborhoods where the sharp peaks are not resolved very well. In Fig.~\ref{fig:resultsunequalspinning}d, the dominant absolute errors are concentrated around the strong-field regions near the punctures, while the error remains comparatively small away from the peaks.

\section{Summary and discussion}
\label{s5}
In this work, we used Physics-Informed Neural Networks (PINNs) to solve the highly nonlinear Hamiltonian constraint equation for binary black hole initial data in numerical relativity for the first time. Our results indicate that using PINN to solve Einstein equation is possible.

To overcome the shortcomings of traditional PINN methods in solving this problem, we introduced an ansatz that incorporates an analytical guiding solution into the network architecture. This strategy provides explicit physical guidance, significantly decreasing the optimization difficulty, and leading to solutions with remarkably high accuracy.

Furthermore, in order to resolve the features near the punctures, we formulated a composite loss optimization framework that can combine the $L_2$ loss with the $\text{soft-}L_\infty$ loss. This framework helps the network better preserve local fidelity around the punctures and, in most cases, reduce the residuals more effectively.

Moreover, to mitigate the extreme scale disparities among different losses, we implemented a loss balancing strategy, so that the losses remain at comparable numerical scales during training, ensuring highly stable convergence across the entire computational domain.

For the hyperparameters used in this work, we considered several representative values in our tests. Specifically, we tested $c=\{0.2,0.25,0.3,0.4,0.5,1.0\}$, $w_\infty=\{0,0.2,0.5,1.0\}$, $\mathrm{lr}=\{1\times 10^{-4},3\times 10^{-4},5\times 10^{-4}\}$, and $\beta=\{2.5,10,20\}$. We did not perform a full grid search over all possible combinations, since it would require a larger computational cost. The final values used in each physical case were selected from these preliminary runs based on training stability and the resulting L2RE relative to the \texttt{TwoPunctures} reference solution. In addition, we observed that the number of layers and the number of neurons per layer also affect the prediction accuracy. Therefore, the method currently relies on manual adjustment of training hyperparameters, which remains a notable limitation of the present work.

For the computational cost, the present PINN implementation is still more expensive than \texttt{TwoPunctures} for a single binary black hole configuration. In our tests, training the PINN for $5000$ steps on a single NVIDIA L40 GPU takes about $30$ minutes, which is approximately $25$ times longer than the corresponding \texttt{TwoPunctures} computation. Nevertheless, the main purpose of this work is to demonstrate the ability of the PINN method for solving the Hamiltonian constraint equation. We are currently developing a parameterized PINN-based initial-data solver, in which the network is trained within a certain range of physical parameters. Such a framework is expected to improve the efficiency and reduce the manual hyperparameter tuning process.

Overall, our results show that, with appropriately tuned hyperparameters for each physical configuration, the network can predict highly accurate solutions for the non-spinning binary black hole systems and relatively accurate solutions for the spinning configuration. In future work, we aim to develop a fully parameterized initial data solver in which the hyperparameters can be adaptively determined from the physical parameters of the system, leading to an automated framework for generating sufficiently accurate binary black hole initial data.

\begin{acknowledgments}
We thank Chenkai Qiao, Yuchen Yang, Tengfei Xu and Zhuoli Ouyang for useful discussions. This work was supported in part by the National Key Research and Development Program of China Grant No. 2021YFC2203001 and in part by the NSFC (No.~12475046).
\end{acknowledgments}

\bibliography{refs}

\end{document}